\documentclass[12pt]{article}

\usepackage{graphicx}
\usepackage{setspace}
\usepackage{amsmath}
\usepackage{latexsym, amssymb}
\usepackage{verbatim}
\newcommand{\bea}{\begin{eqnarray}}  
\newcommand{\eea}{\end{eqnarray}}
\newcommand{\ben}{\begin{enumerate}}
\newcommand{\een}{\end{enumerate}}
\newcommand{\m}[1]{\mathcal{#1}}
\newcommand{\p}{\partial}

\newcommand\pubnumber{SLAC--PUB--15832}
\newcommand\pubdate{\today}
\newcommand\pubblock{\rightline{\begin{tabular}{l} \pubnumber\\
         \pubdate \end{tabular}}}

\def\SLAC{SLAC National Accelerator Laboratory, Menlo Park, CA 94025 USA}
\def\doeack{\footnote{Work supported by the US Department of Energy,
                     contract DE--AC02--76SF00515.}}
\def\Title#1{\begin{center} {\Large #1 } \end{center}}
\def\Author#1{\begin{center}{ \sc #1} \end{center}}
\def\Address#1{\begin{center}{ \it #1} \end{center}}

\begin{document}
\begin{titlepage}
\pubblock

\bigskip
\Title{Three-point current correlation functions as probes of Effective Conformal Theories}
\Author{Kassahun Betre\doeack}
\Address{\SLAC}
\bigskip

\begin{abstract}
The three-point current correlation function in Euclidean spacetime for a strongly coupled system with non-Abelian global symmetry, $\langle J^a_i(x)J^b_j(y)J^c_k(z)\rangle$, is calculated from the weakly coupled AdS dual. The contribution from the first non-renormalizable bulk operator, $(F_{\mu\nu})^3$, is calculated and shown to lead to a polarization structure different from the leading contribution, which comes from the renormalizable $(F_{\mu\nu})^2$ operator. The non-renormalizable correction is suppressed by powers of the cutoff scale $\Lambda$. This suggests a possible experimental probe of the effective description through a measurement of the cutoff scale $\Lambda$ in strongly coupled condensed matter systems.
\end{abstract}
\end{titlepage}

\section{Introduction}
The AdS/CFT correspondence is a powerful tool for computing observables in strongly coupled systems with conformal symmetry by mapping them to weakly coupled dual gravitational theories. However, our ability to exploit the correspondence is limited by our ability to compute in the weakly coupled theory itself. For example, on the bulk AdS side, theories of practical use are not only weakly coupled, but also ``well behaved," in the sense that they are effective theories describing the dynamics of only a few fields below some cutoff scale $\Lambda$. The cutoff scale suppresses non-renormalizable operators generated when fields above the cutoff scale are integrated out.

This leads to the line of enquiry: what is the class of CFTs that we can explore by mapping them to weakly coupled, well-behaved AdS duals? Put another way, what are the necessary and sufficient conditions needed for a CFT to have a weakly coupled, well behaved AdS dual?

Explorations along those lines gave rise to the idea of Effective Conformal Theories (ECT) \cite{Fitzpatrick:2010zm}. The idea of ECTs is that the strongly coupled CFTs that can be described through weakly coupled, effective AdS bulk theories are characterized by two conditions: (1) There is a large dimension gap in the spectrum of the dilatation operator. (2) There is a small parameter that suppresses higher point connected correlation functions. These conditions are naturally satisfied in large-$N$ models where $1/N$ plays the role of the small parameter.

In this paper we explore one consequence of such effective conformal descriptions. Assuming that such an effective description is valid for a strongly coupled condensed matter system with non-Abelian global symmetry, the three-point current correlation function $\langle J^a_i(t_1,x)J^b_j(t_2,y)J^c_k(0)\rangle$ admits a perturbative expansion in the parameter $\Delta = (\Lambda R_{AdS})$. The successive terms in the series carry different polarization structures. In the bulk effective AdS, the dominant contribution to the three point current correlation function comes from the renormalizable (for $d\ge4$) bulk operator $(F_{\mu\nu})^2$. The second contribution comes from a non-renormalizable $(F_{\mu\nu})^3$ operator. In this paper we will refer to these two operators as $F^2$ and $F^3$ respectively. The latter operator is suppressed by the mass scale $\Lambda$. The suppression in the boundary dual is by the parameter $\Delta = \Lambda R_{AdS}$. We will show that generally the $F^3$ operator leads to a different polarization structure for the three-point current correlation function. This difference can be exploited to experimentally measure the expansion parameter $\Delta$ through the framework of ECTs.

The outline of the paper is as follows. An overview of ECTs is given in Section 2. In Section 3, we give the derivation of the contribution of the bulk $F^3$ term to the boundary three-point current correlation function. This contribution is compared to the dominant contribution coming from $F^2$ term, which is computed in \cite{Freedman:1998tz}. Generalizations of the conformal tensors $D_{ijk}(x,y,z)$ and $C_{ijk}(x,y,z)$ used in $d=4$ dimensions in \cite{Freedman:1998tz} is given to general $d>2$ dimensions. In Section 4, we will outline a possible experimental measurement that can be performed to test the validity of ECT for condensed matter systems.

\section{Effective Conformal Thoeries}
We begin with the question, ``what are the necessary and sufficient conditions needed for a CFT to have a weakly coupled, well behaved AdS dual?"
The necessary conditions were first motivated by locality considerations in type IIB string theory on $AdS_5\times S^5$/$\m{N}=4$ SYM. The regime where the 10D supergravity is a good description (i.e., the regime where there is an approimate 10D flat spacetime in the neighborhood of every point) requires the mass of string excitations, of order inverse string length $l_s^{-1}$, to be hierarchically larger than those of the supergravity modes of order inverse AdS length $R_{AdS}^{-1}$ \cite{Aharony:1999ti}. At energies much smaller than $l_s^{-1}$ the theory will look like a local field theory. Since $R_{AdS}=\lambda^{1/4}l_s$, where the 't Hooft coupling $\lambda = g_{YM}^2N$, the condition that $R_{AdS} >> l_s$ implies that the 't Hooft coupling must be large, $\lambda >> 1$. Applying S-duality, which maps type IIB string theory to itself under $g_s \rightarrow g'_s = 1/g_s$, and demanding that string modes remain heavy in the S-dual of the type IIB, we find another condition. Under S-duality, 
	\begin{eqnarray}
		1<< \lambda & = & g_{YM}^2N \\
		 & \underrightarrow{\hbox{\tiny{S-duality}}} & \lambda' = g_{YM}'^2N = \frac{1}{g_{YM}^2}N = \frac{N^2}{\lambda}
	\end{eqnarray}
The requirement that string modes should remain heavy in both sides of the duality is the statement that both $\lambda>>1$ and $\lambda' >> 1$. We find the simultaneous requirements that $\lambda>>1$ and $N^2/\lambda>>1$, which are satisfied for $N^2>>\lambda$, i.e, $N$ very large. But since $R_{AdS}/l_p \sim N^{1/4}$, where $l_{p}$ is the Planck length, $N >> 1$ implies that $R_{AdS} >> l_{p}$ as well. Then we can ignore supergravity quantum corrections and consider classical or tree level supergravity. 

Therefore, the gravitational bulk theory is an effective field theory with a large mass gap between the fields of mass of order $R_{AdS}^{-1}$ and high mass string and quantum gravitational excitations with masses of order $l_s^{-1}$ and $l_{p}^{-1}$ respectively. The effective theory has a perturbative expansion in the inverse mass gaps which suppress non-renormalizable interactions. In particular, gravitational interactions are suppressed by powers of $M_{p}^{-1}$, so we can ignore graviton exchanges. 

In the dual $\m{N}=4$ Super Yang-Mills theory, the large mass gap in the effective AdS translates to a large gap in operator dimensions. Further, the conformal theory has an expansion in $1/N$, since $N$ is large. This is what mirrors the suppression by factors of $M_p^{-1}$ of gravitational interactions in the AdS bulk. The $1/N$ expansion suppresses higher point connected correlation functions compared to two point functions. Based on this result, Heemskerk, Penedones, Polchiniski, and Sully \cite{Heemskerk:2009pn} put forward the conjecture that any CFT with a large-$N$ like expansion and large gap in the operator dimensions has a local bulk dual AdS theory \footnote{ We also need all single trace operators of spin greater than two to have large dimensions since there is no known local bulk theory of particles of spin greater than 2}. The large $N$ - like expansion parameter is needed to suppress higher point connected functions compared to two point ones, which in the bulk dual corresponds to suppression of gravitational interactions.  Fitzpatrick and Kaplan \cite{Fitzpatrick:2012cg} have shown that with the added condition that the Mellin amplitudes of the CFT correlators have an effective theory-type expansion, we obtain the full set of necessary and sufficient conditions for a CFT to have a well behaved weakly coupled bulk AdS dual.

The picture we obtain is that the weakly coupled, well-behaved AdS duals have a double expansion in $l_s^{-1}$, and $l_p^{-1}$. The question is, what do these expansions correspond to on the CFT side? From the above paragraphs it is clear that one of these expansions is a $1/N$ expansion which suppresses higher point connected correlation functions. But what does the expansion in the inverse dimension gap imply? Is there a concept of ``Effective Conformal Theory (ECT)?" that describes the dynamics of operators whose dimension lies below the cutoff dimension? If so, how does such a theory distinguish between ``renormalizable" vs ``non-renormalizable" interactions? What suppresses the ``non-renormalizable" operators (since conformal symmetry means that there are no mass scales)? What conditions set the range of validity for such an effective conformal theory, and where does it break down? 

To address these questions, Fitzpatrick, Katz, Poland and Simmons-Duffin \cite{Fitzpatrick:2010zm} identified these two expansions with those involving a large parameter $N$ and a large dimension gap $\Delta_{gap} = \Delta_{Heavy} - \Delta_{low}$. Such a theory is an effective conformal theory that captures the dynamics of the low-lying spectrum of the dilatation operator. Let $\Delta_{low}$ be the typical dimension of the low-lying operators, and let all other primary operators have dimension above $\Delta_{Heavy}$ which is hierarchically larger. Then there is a perturbative expansion in both $1/\Delta_{Heavy}$ \cite{Sundrum:2011ic} and $1/N$. The $1/N$ suppresses all interactions, and the $1/\Delta_{Heavy}$ suppresses higher dimensional operators in the OPE. 

There is a direct parallel with effective quantum field theories. In that familiar context, there is an expansion in the small coupling constant of the effective QFT in addition to an expansion in $1/M$, where $M$ is the scale where the effective QFT begins to break down. Analogously, in effective CFTs, the large $N$ (playing the role of the small coupling constant) ensures that connected pieces of higher point correlation functions are suppressed compared to two-point functions, whereas the small $\Delta^{-1}$ (playing the role of small $M^{-1}$ in QFTs) suppresses contributions of higher dimensional operators to the correlation function. 

The schematic picture obtained is therefore the following. The dilatation operator of the CFT has a perturbative expansions in both $1/N$ and $1/\Delta_{Heavy}$:
	\bea
		D^{eff} = D^0 + \frac{1}{N}\left( V^{(1)} + \frac{1}{\Delta_{Heavy}}V^{(2)} + \dots \right) + \mathcal{O}(\frac{1}{N^2}),
	\eea
where $D^0$ is the mean field dilatation operator and $V^{(1)}, V^{(2)}, \dots$ are perturbations of the dilatation that preserve conformal symmetry. 

The next question is then, ``what sets the range of validity of the effective description?" The answer is again analogous to the situation in effective field theories where imposing perturbative unitarity on the Hamiltonian sets the range of validity of the effective theory. In our case, perturbative unitarity is imposed on the dilatation operator  \cite{Fitzpatrick:2010zm}. Assume $\m{O}$ is the only single trace primary operator below the cutoff dimension $\Delta_{Heavy}$. Then the low dimensional spectrum of the dilatation consists of double trace primary operators of the type $\m{O}_{n,l}=\m{O}(\p^2)^n(\p)^l\m{O}$. These operators receive an order $1/N$ correction to their dimension coming from the $V^{(1)}$ term; $\Delta_{n,l} = 2\Delta + 2n+l+\frac{1}{N}\gamma(n,l)$. Imposing perturbative unitarity gives a bound $|\gamma(n,l)| < 4$ on the anomalous dimension $\gamma(n,l)$. However, operators $V^{(1)}$ dual to bulk interactions of mass (or scaling dimension) $\Lambda^p$ (hence forth refered to as ``non-renormalizable" operators) lead to growth in $\gamma(n,l)$ as $n^{p-(d+1)}$ \cite{Heemskerk:2009pn, Fitzpatrick:2010zm}. Even though $\gamma(n,l)$ is an $O(1/N)$ correction, it leads to violation of the unitarity bound for $p>d+1$ and sufficiently large $n$ no matter how small $1/N$ may be. As $n$ approaches $\Delta_{Heavy}$, the new operators must be integrated in to moderate the growth of $\gamma(n,l)$ and restore unitarity. This will indeed be the case if the non-renormalizable operators $V$ of dimension $p$ are suppressed by $\Delta_{Heavy}^{p-(d+1)}$. In this case, $\gamma(n,l)$ grows as $\left(n/\Delta_{Heavy}\right)^{p-(d+1)}$, the unitarity bound is satisfied as long as $n < \Delta_{Heavy}$, and the ECT breaks down when $n \sim \Delta_{Heavy}$.

This idea to use perturbative unitarity as the condition to set the range of validity of the effective description was suggested by the authors of \cite{Fitzpatrick:2010zm} as a solution to the observation made in \cite{Hofman:2008ar} that in correlation functions involving conserved currents, only certain polarization structures, those arising from the lowest dimension bulk operators appear. Demanding perturbative unitarity on all operators below the cutoff dimension $\Delta < \Delta_{Heavy}$ translates to demanding that the scale suppressing non-renormalizable operators in the bulk satisfy $\Lambda > \left(\Delta_{Heavy}/R_{AdS}\right)$. By explicitly computing the contribution of the bulk operator $F^3$ to the three-point current correlation function, we will show that, in addition to giving a polarization structure different from that of the $F^2$, the contribution is suppressed by the appropriate power of the cutoff dimension $\Delta_{Heavy}$.  

\section{Three-point Current Correlation Function}
Armed with the above perturbative expansion, we can compute the three-point current correlation function resulting from the operator $F^3$ and compare the result to the contribution of the $F^2$ operator. It is important to note here that the system remains  conformally invariant in the presence of the non-renormalizable $F^3$ operator. This is guaranteed by the fact that in the bulk $AdS$ the operator is invariant under the $AdS$ isometry. The theory we are describing thus models movement along a line of second order phase transition of a system with non-Abelian global symmetry. The movement is parameterized by $\Lambda R_{AdS}$. 

We begin with the bulk action
	\bea
	\label{bulkaction}
		S &=& \frac{1}{g^2_{SG}} \int d^{d+1} x \sqrt{g} \left(  \frac{1}{2}F^2 + \Lambda^{-p}F^3 \right).
	\eea
$g_{SG}$ is the gauge coupling constant for the bulk $AdS$ Yang-Mills theory. $\Lambda$ has mass dimension $+1$. The explicit form of the operator $F^3$ that we will be using is 
	\bea
		F^3 = f^{abc}F_{\mu\alpha}^aF_{\nu\beta}^bF_{\rho\gamma}^c g^{\alpha \nu}g^{\beta \rho}g^{\gamma \mu}.
	\eea
Note however that we do not need non-Abelian global symmetry to get $F^3$ term. If there are three $U(1)$ global currents in the boundary CFT, we will get bulk interaction terms of the form $F_{\mu\alpha}F^{\alpha\nu}F_{\nu}^{\mu}$. However, in this case there are no renormalizable bulk interactions that contribute to the three-point current correlation function, the first non-vanishing contribution being the $F^3$. 

Through out this paper we will be working in Euclidean $AdS$ and have rescaled the gauge fields so that $A_{\mu}\to (i/g_{SG})A_{\mu}, F_{\mu\nu} \to (i/g_{SG})F_{\mu\nu}.$ Further, the gauge group generators have the commutation relation $[T^a,T^b] = f^{abc}T^c$. With these modifications we have $$F_{\mu\nu} = \p_{\mu}A_{\nu} - \p_{\nu}A_{\mu} + [A_{\mu}, A_{\nu}].$$
Dimensional analysis gives the following mass dimensions:
	\bea
		\left[ g_{SG} \right] &=&  \frac{3-d}{2}\nonumber\\
		\left [F \right] &=& 2\nonumber\\
		 p &=&  2 \nonumber
	\eea
Let us write the action as $S = S_2 + S_3$ where $S_2$ is the $F^2$ integral and $S_3$ is the $F^3$. We study contributions to the three-point current correlation function $\left \langle J_i^a(x)J_j^b(y)J_k^c(z) \right \rangle$ coming from each of the actions $S_2$ and $S_3$. $i,j,k$ are $d-$dimensional Euclidean spacetime indices and $a,b,c$ label global current indices. The points $x,y,z \in \mathbb{R}^{d}$ are points in $d$-dimensional Euclidean spacetime. In this paper we adapt the notation of \cite{Freedman:1998tz}, where the contribution of $S_2$ has been computed. Let us first begin with a review of the conformal structures of the two and three-point current correlation functions.

\subsection{Review of conformal structures}
The two-point current correlation function in $d-$dimensions is fully determined by conformal invariance up to a normalization constant. It is given by
	\bea
	\label{2point}
		\left \langle J_i^a(x)J_i^b(y)\right \rangle	&=& B\delta^{ab}\frac{2(d-1)(d-2)}{(2\pi)^d}\frac{J_{ij}(x-y)}{|x-y|^{2(d-1)}}.
	\eea
$B$ is a positive constant and $$J_{ij}(x) = \delta_{ij} - 2\frac{x_ix_j}{x^2}.$$
The coefficient $B$ is computed from the bulk $F^2$ term in \cite{Freedman:1998tz},
	\bea
		B = \frac{1}{g^2_{SG}}\frac{2^{d-2}\pi^{\frac{d}{2}}\Gamma(d)}{(d-1)\Gamma(\frac{d}{2})}.
	\eea

The three-point current correlation function is also determined completely by conformal symmetry up to two constants. In $d=4$ dimensions, the normal parity three-point function is given as the superposition of two permutation-odd conformal tensor structures, $D^{sym}_{ijk}, C^{sym}_{ijk}$ \cite{Freedman:1992tz}. 
	\bea
		\left \langle J_i^a(x)J_j^b(y)J_k^c(z) \right \rangle_{+} &=& f^{abc}\left( k_1 D_{ijk}^{sym} + k_2 C_{ijk}^{sym}\right)
	\eea
where 
	\bea
		D^{sym}_{ijk}(x,y,z) &=& D_{ijk}(x,y,z)+D_{ijk}(z,x,y)+D_{ijk}(y,z,x)\\
		C^{sym}_{ijk}(x,y,z) &=& C_{ijk}(x,y,z)+C_{ijk}(z,x,y)+C_{ijk}(y,z,x)\\
	\eea
The tensors $D_{ijk}(x,y,z),$ and $C_{ijk}(x,y,z)$ are given by
	\bea
		D_{ijk}(x,y,z) &=& \frac{1}{(x-y)^2(y-z)^2(z-x)^2}\frac{\p}{\p x_{i}}\frac{\p}{\p y_{j}}\ln\left( (x-y)^2\right)\frac{\p}{\p z_{k}}\ln\left( \frac{(x-z)^2}{(y-z)^2}\right)\\
		&=& \frac{4}{(x-y)^2(y-z)^2(z-x)^2}J_{ij}(x-y)\frac{\tilde{t}_{k}}{(x-y)^2}
	\eea
	\bea
		C_{ijk}(x,y,z) &=& \frac{1}{(x-y)^4}\frac{\p}{\p x_{i}}\frac{\p}{\p z_{l}}\ln\left( (x-z)^2\right)\frac{\p}{\p y_{j}}\frac{\p}{\p z_{l}}\ln\left( (y-z)^2\right)\frac{\p}{\p z_{k}}\ln\left( \frac{(x-z)^2}{(y-z)^2}\right)\\
		&=& \frac{-8}{(x-y)^2(y-z)^2(z-x)^2}J_{il}(x-z)J_{jl}(y-z)\frac{\tilde{t}_{k}}{(x-y)^2},
	\eea
where,
	\bea
		\tilde{t}_k = \frac{(x-z)_k}{(x-z)^2}-\frac{(y-z)_k}{(y-z)^2},\quad t_k = \frac{(y-x)_k}{(y-x)^2}-\frac{(z-x)_k}{(z-x)^2},\quad\hat{t}_k = \frac{(z-y)_k}{(z-y)^2}-\frac{(x-y)_k}{(x-y)^2}.
	\eea
The vectors $t$ and $\hat{t}$ are introduced here for later convenience since they appear in the symmetric sums of $D_{ijk}$, and $C_{ijk}$.
In $d=4, C^{sym}_{ijk}$ satisfies $\frac{\p}{\p z_k}C^{sym}_{ijk} = 0$ everywhere, whereas $D^{sym}_{ijk}$ has terms proportional to $\delta^4(z-x)$ and $\delta^4(z-y)$. Therefore, the Ward identity in $d=4$ relates the coefficient  $k_1$ to the coefficient $B$ in (\ref{2point}) as 
	\bea
		\label{d4wardid}
		k_1 = \frac{B}{16\pi^6}.
	\eea
 The coefficient $k_2$ is undetermined.

The contribution to the three-point function coming from the bulk action $S_2$ is calculated for general $d$ in \cite{Freedman:1998tz}.  
	\bea
	\label{fijkS2}
		\left\langle J^a_i(x)J^b_j(y)J^c_k(z)\right\rangle^{S_2}
		&=&\frac{f^{abc}}{2g^2_{SG}\pi^4} 2\left[ \mathcal{F}^{(2)}_{ijk}(x,y,z)+\mathcal{F}^{(2)}_{kij}(z,x,y)+							\mathcal{F}^{(2)}_{jki}(y,z,x)\right],\\
			\mathcal{F}^{(2)}_{ijk}(x,y,z) &=& -\kappa \frac{J_{jl}(y-x)}{|y-x|^{2(d-1)}}\frac{J_{km}(z-x)}{|z-x|^{2(d-1)}}\nonumber\\
		&&\times\frac{1}{|t|^d}\left[ \delta_{lm}t_i+(d-1)\delta_{il}t_m+(d-1)\delta_{im}t_l-d\frac{t_it_lt_m}{|t|^2}\right]\nonumber
	\eea
where,
	\bea
		\kappa 
		&=& \pi^{d/2}(C^d)^3\frac{(d-2)}{(d-1)}\frac{\big[ \Gamma(\frac{d}{2})\big]^3}{\big[ \Gamma(d)\big]^2}, \quad C^d = \frac{\Gamma(d)}{2\pi^{d/2}\Gamma(\frac{d}{2})}\nonumber
	\eea	
In terms of the conformal tensors $D^{sym}_{ijk}, C^{sym}_{ijk}$, the above result takes the elegant form
	\bea
	\label{jjjS2}
		\left\langle J^a_i(x)J^b_j(y)J^c_k(z)\right\rangle^{S_2}
		&=&\frac{f^{abc}}{2g^2_{SG}\pi^4}\left( D^{sym}_{ijk}-\frac{1}{8}C^{sym}_{ijk}\right).		
	\eea
	
Let us digress here to comment on the comparison between this bulk result for the lowest renormalizable operator $F^2$ in $d=4,$ with the 1-loop exact two and three-point correlation function in the boundary $\mathcal{N}=4$ super-Yang-Mills theory. With the replacement $4\pi/N \rightarrow g_{SG}$ we find that both the two-point and three-point correlation functions agree exactly. 

In the two-point function, from the boundary super-Yang-Mills perspective, there are no higher order corrections than the 1-loop result because of powerful non-renormalization theorems \cite{Anselmi:1997am}. But on the bulk side, we would expect that bulk operators of the form 
	\bea
		\label{nonrenorm}
		\sum_n \frac{1}{g^2_{SG}\Lambda^{2n}}\left( (\p_{\rho}\p^{\rho})^n F_{\mu\nu}F^{\mu\nu} + \p_{\mu_1}\p_{\mu_2}\dots\p_{\mu_n}F_{\mu\nu}\p^{\mu_1}\p^{\mu_2}\dots\p^{\mu_n}F^{\mu\nu} \right)
	\eea
would lead to contributions. These are all operators of the same order in $1/N$ expansion compared to the leading $F^2$ term. Since supergravity is an effective theory that starts to break down when we get near the string scale, we will in fact have the above non-renormalizable operators below the string scale. It must then be the case that the $\m{N}=1$ supergravity of the $AdS_5\times S^5$ is responsible for the vanishing all such contributions.

If we remove supersymmetry from both sides of the duality, non-renormalizable operators of the form (\ref{nonrenorm}) will lead to corrections to $B$. Similar corrections arise for the three-point function. The claim of \cite{Fitzpatrick:2010zm} is that effective bulk theories where non-renormalizable operators of the form (\ref{nonrenorm}) are suppressed by appropriate mass scales are dual to effective conformal theories where perturbative unitarity is imposed on the dilatation operator. By computing the contribution of the $S_3$ action to the three-point correlation function, we will demonstrate that contributions to $k_1$ and $k_2$ coming from the non-renormalizable bulk operator $F^3$ will be suppressed by $\Delta_{gap}^2 = (R_{AdS}\Lambda_{cutoff})^2$ as required by perturbative unitarity on the dilatation on the CFT side. In addition, we will see that the contribution of the $F^3$ operator has different polarization structure, which could be exploited to experimentally measure the suppression parameter $\Delta$.  

\subsection{Generalization in $d>2$} 
In $d > 2$ dimensions, the symmetric tensor $J_{ij}$ which appears in the two-point function in (\ref{2point}) remains the same since it comes from general requirements of covariance under the conformal algebra \cite{Osborn:1993cr}. 
The tensors $D_{ijk}(x,y,z),$ and $C_{ijk}(x,y,z)$ can be generalize as follows. 
	\bea
		D_{ijk}(x,y,z) &=& \frac{1}{\big( |x-y||y-z||z-x|\big)^{d-2}}\frac{\p}{\p x_{i}}\frac{\p}{\p y_{j}}\ln\left( |x-y|^{d-2}\right)\frac{\p}{\p z_{k}}\ln\left( \frac{|x-z|^{d-2}}{|y-z|^{d-2}}\right)\\
		&=& \frac{(d-2)^2}{\big(|x-y||y-z||z-x|\big)^{d-2}}J_{ij}(x-y)\frac{\tilde{t}_{k}}{|x-y|^2}\\
	\eea
	\bea
		C_{ijk}(x,y,z) &=& \frac{1}{|x-y|^d}\frac{\p}{\p x_{i}}\frac{\p}{\p z_{l}}\ln\left( |x-z|^{d-2}\right)\frac{\p}{\p y_{j}}\frac{\p}{\p z_{l}}\ln\left( |y-z|^{d-2}\right)\frac{\p}{\p z_{k}}\ln\left( \frac{|x-z|^{d-2}}{|y-z|^{d-2}}\right)\\
		&=& \frac{-(d-2)^3}{\big(|x-y||y-z||z-x|\big)^{d-2}}J_{il}(x-z)J_{jl}(y-z)\frac{\tilde{t}_{k}}{|x-y|^2},
	\eea
The symmetric sums of the tensors, $D^{sym}_{ijk}, C^{sym}_{ijk}$ have the following property:
	\bea
		\frac{\p}{\p z_{k}}D^{sym}_{ijk}&=&(d-2)^2 S_d\left(\frac{d+2}{d}\right)\frac{J_{ij}(x-y)}{|x-y|^{2(d-1)}}\Big(\delta^d(z-y) - \delta^d(z-x) \Big)\nonumber\\
		\frac{\p}{\p z_{k}}C^{sym}_{ijk}&=&-(d-2)^3S_d\left( \frac{d-4}{d}\right)\frac{J_{ij}(x-y)}{|x-y|^{2(d-1)}}\Big(\delta^d(z-y) - \delta^d(z-x) \Big),
	\eea
where, $$S_d = \frac{2\pi^{\frac{d}{2}}}{\Gamma(\frac{d}{2})}.$$ We have used the following formulae to derive the above result:
	\bea
		\lim_{x\to 0} \frac{x_ix_j}{x^2} = \frac{1}{d}\delta_{ij}, \qquad \lim_{z\to x}\frac{\p}{\p z_k}\left( \frac{(z-x)_k}{|z-x|^d}\right) = S_d \delta^d(z-x).
	\eea
The Ward identity in $d-$dimensions relates one linear combination of $k_1$ and $k_2$ to $B$. 
	\bea
	\label{ddwardid}
		B &=& \frac{(2\pi)^dS_d}{2}\frac{(d-2)}{(d-1)} \left( \frac{(d+2)}{d}k_1 -\frac{(d-2)(d-4)}{d}k_2\right).
	\eea
In $d=4$ we recover (\ref{d4wardid}).

To compare the contribution of the $F^3$ operator to the three-point function with that coming from the $F^2$ operator in general $d > 2$ dimensions, it is helpful to find an expression to (\ref{fijkS2}) analogous to (\ref{jjjS2}) for general $d>2$ dimensions. This can be achieved using the formulae
	\bea
		J_{km}(z-x)t_m &=& -\frac{(y-z)^2}{(y-x)^2}\tilde{t}_k, \; \text{and}\nonumber\\
		J_{jl}(y-x)t_l &=& -\frac{(z-y)^2}{(z-x)^2}\hat{t}
	\eea
We then find 
	\bea
		\left\langle J^a_i(x)J^b_j(y)J^c_k(z)\right\rangle^{S_2}
		&=&\frac{f^{abc}\kappa}{2g^2_{SG}}\frac{(3d-4)}{(d-2)^2}\left( D^{sym}_{ijk}-\frac{1}{(3d-4)}C^{sym}_{ijk}\right)
	\eea


\subsection{Contribution of the $F^3$ operator}
From the AdS/CFT ansatz for correlation functions \cite{Witten:1998qj}, we have 
	\bea
		\left \langle \exp \int J^a_iA_0^{ai} \right \rangle_{CFT} &=& Z_S(A_0)
	\eea
where $Z_S(A_0)$ is the bulk path integral for the gauge field $A(x_0,x)$ expressed in terms of the boundary value $A_0(x)$. In the limit where the bulk gravitational theory is weakly coupled, the path integral is approximately the classical path integral,
	\[
		Z_S(A_0)  \simeq  \exp (-I_s(A_0)),
	\]
where $I_s(A_0)$ is the action expressed in terms of the boundary value of the field $A$ at boundary coordinates, $x,y,z$. In the following, Latin indices i,j,k run from 1 to $d$, and Greek letters $\mu,\nu$ run from 0 to $d$, where 0 is the extra $AdS$ coordinate.

We are interested in the connected three point correlator,
	\bea
	\label{connected3pt}
 		\left \langle J_i^a(x)J_j^b(y)J_k^c(z) \right \rangle_{connected}
		&=& 
		\frac{\delta}{\delta A_0^{ai}(x)} \frac{\delta}{\delta A_0^{bj}(y)} \frac{\delta}{\delta A_0^{ck}(z)} \log(Z_S(A_0))\nonumber\\
		&=&
		\frac{\delta}{\delta A_0^{ai}(x)} \frac{\delta}{\delta A_0^{bj}(y)} \frac{\delta}{\delta A_0^{ck}(z)}(-I_s(A_0))			\arrowvert_{A_0 = 0}
	\eea

To compute the contribution of the $F^3$ operator, we begin by expressing the $S_3$ part of the action in terms of the boundary value of the gauge field and the boundary-to-bulk Greens function $G^{ab}_{\mu i}(w_0,x;0,\tilde{x})$, where $x,\tilde{x}$ are the $d$-dimensional boundary coordinates and $w_0$ is the perpendicular bulk coordinate. 
	\bea
		A_{\mu}^a(w_0, \tilde{x})&=&
		\int d^dxG^{ab}_{\mu i}(w_0,\tilde{x};0,x)A_0^{ib}(0,x), \hspace{3mm} \hbox{where} 
		G^{ab}_{\mu i} = G_{\mu i}\delta^{ab} \hspace{3mm} \hbox{and so}\nonumber\\
		A_{\mu}^a(w_0, \tilde{x})&=&
		\int d^dxG_{\mu i}(w_0,\tilde{x};0,x)A_0^{ia}(0,x)
	\eea

Plugging this into the $S_3$ part of the bulk action in (\ref{bulkaction}) and evaluating (\ref{connected3pt}) we find the following expression.

	\bea
	\label{JJJ}
	 	\left \langle J_i^a(x)J_j^b(y)J_k^c(z) \right \rangle^{S_3}_{connected}
	 	&=&\frac{\delta}{\delta A_0^{ai}(x)}\frac{\delta}{\delta A_0^{bj}(y)} \frac{\delta}{\delta A_0^{ck}(z)}
		(-S_3)\arrowvert_{A_0 = 0}\nonumber\\
	 	&=&\frac{1}{\Lambda^p g^2_{SG}}2f^{abc}[\mathcal{F}^{(3)}_{ijk}+\mathcal{F}^{(3)}_{jki}+\mathcal{F}^{(3)}_{kij}],
	\eea
where
	\bea
		\label{f3ijk}
		\mathcal{F}^{(3)}_{ijk}
		&=&\int d^{d+1}w\sqrt{g}\hspace{1mm}\p_{[ \mu}G_{\alpha ]i}(w,x)\p_{[ \nu}G_{\beta ]j}(w,y)
		\p_{[ \rho}G_{\gamma ]k}(w,z)g^{\alpha \nu}g^{\beta \rho}g^{\gamma \mu}.
	\eea

We evaluate $\mathcal{F}^{(3)}_{ijk}$ in Euclidean $AdS$, in the parameterization of $AdS$ as the Lobachevsky upper half space with the metric 
	\bea
	\label{metric}
		ds^2 = \frac{R_{AdS}^2}{w_0^2}\left( dw_0^2+\sum_{\mu = 1}^d dx_{\mu}^2\right).
	\eea
We set $R_{AdS} = 1$ in the following computation and restore it in the final answer by dimensional analysis.

The boundary-to-bulk propagator of the gauge field from the boundary point $x^{\mu}=(0,x)^{\mu}$ to the bulk point $w^{\mu}=(w_0,\tilde{x})^{\mu}$ is given explicitly in \cite{Freedman:1998tz}
	\bea
	\label{gaugepropagator}
		G_{\mu i}(w_0,\tilde{x};0,x)
		&=&C^d\frac{w_0^{d-2}}{[w_0^2+(\tilde{x}-x)^2]^{d-1}}J_{\mu i}(w-x).
	\eea

We will use the technique described by Freedman, Mathur, Matusis, and Rastelli \cite{Freedman:1998tz} to evaluate $\mathcal{F}^{(3)}_{ijk}.$ Their technique takes advantage of the fact that the Green function has translation invariance in the boundary coordinates. 
	\[
		\left \langle J_i^a(x)J_j^b(y)J_k^c(z) \right \rangle = \left \langle J_i^a(0)J_j^b(y-x)
		J_k^c(z-x) \right \rangle
	\]
Evaluating $\langle J(0)J(y-x)J(z-x) \rangle$ is easier because there are only two terms in the denominator of (\ref{f3ijk}). We begin by calculating $\left \langle J_i^a(0)J_j^b(y)J_k^c(z) \right \rangle.$ Using the metric (\ref{metric}) in the formula for $\mathcal{F}_{ijk}$ we find, 
	\bea
	\label{Fijk}	
		\mathcal{F}^{(3)}_{ijk}
		&=&\int d^dx' dw_0\frac{w_0^6}{w_0^{d+1}}\p_{[ \mu}G_{\nu ]i}(x',0)\p_{[ \nu}G_{\rho ]j}(x',y)\p_{[ \rho}G_{\mu ]k}
		(x',z)
	\eea
To simplify the above integral further we will take advantage of the inversion isometry of the $AdS$ metric. The transformation 
	\bea
	\label{inversion}
		w_0 = \frac{w'_0}{w'^2_0+x'^2}, \quad x^{\mu} = \frac{x'^{\mu}}{w'^2_0+x'^2} 
	\eea
on the AdS coordinates leaves the metric (\ref{metric}) invariant. On the other hand, such a transformation acts as conformal isometry on the boundary coordinates;  the flat boundary metric $ds^2 = \sum_i dx^idx^i \rightarrow \frac{1}{|x|^4}\sum_i dx^idx^i$ under 
	 \bea
	 \label{boundaryinversion}
	 	x^i &=& \frac{x'^i}{x'^2}.
	 \eea 
The Jacobian of the inversion transformation inherits the tensor structure of $J_{\mu\nu}$
	\bea
		\frac{\p w'_{\mu}}{\p w_{\nu}}&=&w'^2\left( \delta_{\mu \nu}-2\frac{w'_{\mu}w'_{\nu}}{w'^2}\right)\\
		&=&w'^2J_{\mu \nu}(w') = \frac{1}{w^2}J_{\mu\nu}(w)
	\eea
$J_{\mu \nu}$ satisfies the following identities: 
	\bea
	\label{jmunuproperties}
		J_{\mu \nu}(w-u)
		&=& J_{\mu \rho}(w')J_{\rho \sigma}(w'-u')J_{\sigma \nu}(u')\\
		J_{\mu \nu}(w)J_{\nu \rho}(w)
		&=&\delta_{\mu \rho}
	\eea
Using these identities and explicit formula for $G_{\mu \nu}$ we can show that it transforms as a covariant rank 2 tensor with scaling dimension $d-2$ under the simultaneous bulk and boundary inversions.
	\bea
		G_{\mu i}(w_0,\tilde{x};0,x) 
		&=&C^d\frac{1}{w_0}\left(\frac{w_0}{w_0^2+(\tilde{x}-x)^2}\right)^{d-1}J_{\mu i}(w-x)\nonumber\\
		&=&C^d \frac{w'^2}{w'_0}\left(\frac{w'_0}{{w'_0}^2+(\tilde{x}'-x')^2}\right)^{d-1}|x'|^{2(d-1)}J_{\mu \rho}(w')J_{\rho k}(w'-x')J_{ki}(x')\nonumber\\
		&=&w'^2J_{\mu \rho}(w')|x'|^2J_{ki}(x') |x'|^{2(d-2)}G_{\mu i}(w',x')\nonumber\\
		&=&\frac{\p w'_{\nu}}{\p w_{\mu}}\frac{\p x'_k}{\p x_i}|x'|^{2(d-2)}G_{\nu k}(w', x')\nonumber\\
		&=& \frac{\p w'_{\nu}}{\p w_{\mu}}\frac{\p x'_k}{\p x_i}G'_{\nu k}(w', x').
	\eea
In the second line, and $w'^{\mu} =(w'_0,\tilde{x}')^{\mu} $. Similarly, $\p_{[\mu}G_{\nu ]i}(w,x)$ transforms covariantly as
	\bea
		\p_{[\mu}G_{\nu ]i}(w,x)
		&=&w'^2J_{\mu \alpha}(w')w'^2J_{\nu \beta}(w')|x'|^2J_{ik}(x')|x'|^{2(d-2)}\p'_{[\alpha}G_{\beta]k}
			(w',x'),\hspace{2mm}\hbox{where}\\
		\p'
		&=&\frac{\p}{\p w'}\nonumber.
	\eea
When we set $x$ to zero and do an inversion transformation, we find the following simpler forms
	\bea
		G_{\mu i}(w, 0)
		&=& C^d (w'_0)^{d-2}w'^2J_{\mu i}(w')\\
		\p_{[\mu}G_{\nu ]i}(w, 0) 
		&=&(d-2)C^d(w'_0)^{d-3}(w')^4J_{0[\mu}(w')J_{\nu]i}(w'),\hspace{2mm}
	\eea
Applying the inversion on (\ref{Fijk}) and simplifying, we find
	\bea
	\label{Fijk2}
		\mathcal{F}^{(3)}_{ijk}(0,y,z)&=&(d-2)^3(C^d)^3|y|^{2(d-1)}J_{aj}(y')|z|^{2(d-1)}J_{bk}(z) \nonumber\\
		&& \int d^dw'dw'_0\frac{(w'_0)^{2d-4}}{[w'^2_0+(\tilde{x}'-y')^2]^{d-1}[w'^2_0+(\tilde{x}'-z')^2]^{d-1}}\nonumber\\
		&&\hspace{1.7cm} \bigg(J_{0[i}(w'-y')J_{\gamma]a}(w'-y')J_{0[\gamma}(w'-z')J_{0]b}(w'-z')\nonumber\\
		&&\hspace{1.8cm}   +J_{0[\gamma}(w'-y')J_{0]a}(w'-y')J_{0[\gamma}(w'-z')J_{i]b}(w'-z')\bigg)
	\eea
After performing the integral and expressing the result in terms of the tensors $D_{ijk}, C_{ijk}$, we find the following simple form:
	\bea
		\mathcal{F}^{(3)}_{ijk}(x,y,z) &=& -\frac{\kappa d}{2}\Big(D_{jki}(y,z,x) + \frac{1}{d}C_{jki}(y,z,x)\Big), \nonumber
	\eea
The intermediate steps are included in the appendix.  The symmetric sum then becomes
	\bea
		\mathcal{F}^{(3)sym}_{ijk}(x, y, z) 
		&=&-\frac{\kappa d}{2}\Big(D^{sym}_{ijk}(y,z,x) + \frac{1}{d}C^{sym}_{ijk}(y,z,x)\Big). 	
	\eea
For comparison, the contribution of the $F^2$ operator to three-point current correlation function, given in (\ref{fijkS2}) is
	\bea
		\mathcal{F}^{(2)sym}_{ijk}	
		&=&\frac{\kappa(3d-4)}{2(d-2)^2}\left( D^{sym}_{ijk}(x,y,z)-\frac{1}{(3d-4)}C^{sym}_{ijk}(x,y,z)\right).
	\eea
As expected, the polarization structure resulting from the $F^3$ operator is different from the $F^2$ contribution.

After restoring the correct factor of $R_{AdS}$ by dimensional analysis, and letting $R_{AdS}\Lambda = \Delta$, the three-point current contributions of each of the operators $F^2$ and $F^3$ are
	\bea
		\left\langle J^a_i(x)J^b_j(y)J^c_k(z)\right\rangle^{S_2}
		&=&f^{abc}\kappa\left(\frac{(R_{AdS})^{d-3}}{g^2_{SG}}\right)\left(\frac{(3d-4)}{2(d-2)^2}\right)\left( D^{sym}_{ijk}(x,y,z)-\frac{1}{(3d-4)}C^{sym}_{ijk}(x,y,z)\right)\nonumber\\
		\left\langle J^a_i(x)J^b_j(y)J^c_k(z)\right\rangle^{S_3}
		&=&-f^{abc}\kappa  \left(\frac{(R_{AdS})^{d-3}}{\Delta^2g_{SG}^2}\right)d\Big(D^{sym}_{ijk}(y,z,x) + \frac{1}{d}C^{sym}_{ijk}(y,z,x)\Big).	
	\eea
 The sum of the two contributions is, 
	\bea
		\left\langle J^a_i(x)J^b_j(y)J^c_k(z)\right\rangle^{S_2+S_3}
		=f^{abc}\kappa\left(\frac{ (R_{AdS})^{d-3}}{g_{SG}^2}\right)\left( \frac{3d-4}{2(d-2)^2}\right)
		\hspace{-7mm}&&\Bigg[\bigg(1- \frac{2d(d-2)^2}{(3d-4)\Delta^2}\bigg)D^{sym}_{ijk}\nonumber\\
		&&-\frac{1}{3d-4}\bigg(1+\frac{2(d-2)^2}{\Delta^2} \bigg)C^{sym}_{ijk}\Bigg].
	\eea
In particular, for $d=3$,
	\bea
		\mathcal{F}^{(3)sym}_{ijk}
		&=&-\frac{1}{2^9}\left( D^{sym}_{ijk}+\frac{1}{3}C^{sym}_{ijk}\right)\\
		\mathcal{F}^{(2)sym}_{ijk}
		&=&\frac{5}{2^{10}}\left( D^{sym}_{ijk}-\frac{1}{5}C^{sym}_{ijk}\right)\nonumber
	\eea
	\bea
		\left\langle J^a_i(x)J^b_j(y)J^c_k(z)\right\rangle
		&=&f^{abc}\left(\frac{5}{2^{10}g_{SG}^2}\right)\left(\big(1- \frac{6}{5\Delta^2}\big)D^{sym}_{ijk}-\frac{1}{5}\big(1+\frac{2}{\Delta^2} \big)C^{sym}_{ijk}\right).
	\eea

In $d=4$, the combined three-point current correlation function is
	\bea
		\mathcal{F}^{(3)sym}_{ijk}
		&=&-\frac{1}{\pi^4}\left( D^{sym}_{ijk}+\frac{1}{4}C^{sym}_{ijk}\right)\\
		\mathcal{F}^{(2)sym}_{ijk}
		&=&\frac{1}{2\pi^4}\left( D^{sym}_{ijk}-\frac{1}{8}C^{sym}_{ijk}\right)\nonumber
	\eea
	\bea
		\left\langle J^a_i(x)J^b_j(y)J^c_k(z)\right\rangle
		&=&f^{abc}\left(\frac{ R_{AdS}}{2\pi^4g_{SG}^2}\right)\left(\big(1- \frac{1}{4\Delta^2}\big)D^{sym}_{ijk} -\frac{1}{8}\big(1+\frac{8}{\Delta^2} \big)C^{sym}_{ijk}\right).		
	\eea


These results give the two lowest order results to the three-point current correlation function in the $1/\Delta$ expansion and leading order in $1/N$ expansion. The first $\mathcal{O}(1/\Delta^2)$ correction to the three-point current correlation function comes from the non-renormalizable $F^3$ operator. 

\section{Physical measurement}
Measuring the three-point spin-current in condensed matter systems directly is near impossible through existing technologies. However, measurements that look for non-linear Ohm's-law type effects in induced spin-currents contain data about the three-point current correlation function. In the presence of an external field $\vec{E}$ the induced current will take the form,
	\bea
		J_k^c &=& \sigma^{ac}_{ik}E_a^i + d^{abc}_{ijk}E_a^iE_b^j + \mathcal{O}(E^3),
	\eea
With $a,b,c$ indices of global currents, and $i,j,k$ indices of $d-$dimensional Euclidean spacetime coordinates. $\sigma_{ij}^{ab}$ and $d_{ijk}^{abc}$ are the 2 and 3-rank conductivity tensors. The fact that the two operators lead to different polarization structures will be exploited. Consider the special points
	\bea
		z &=& (0,0,0,...,0)\nonumber\\
		x &=& (\tau, r, 0, ...,0)\nonumber\\
		y &=& (\tau, -r, 0,...,0).
	\eea
The $i=j=k$ component of the tensor $D^{sym}_{ijk}$ automatically vanishes, whereas the $ijk=122$ component of $D^{sym}_{122}$ is just a rescaling of $C^{sym}_{ijk}$. However, the $ijk=112$ component of the symmetric tensors $D^{sym}_{112}$ and $C^{sym}_{112}$ are linearly independent, and take the values
	\bea
		D^{sym}_{112}&=& \frac{(d-2)^2}{\Big[2r(\tau^2+r^2)\Big]^{(d-1)}}\left(\frac{r^4-\tau^4+8\tau^2r^2}{(\tau^2+r^2)^2} \right),\nonumber\\
		C^{sym}_{112}&=& - \frac{(d-2)^3}{\Big[2r(\tau^2+r^2)\Big]^{(d-1)}} \left( 1-\frac{16\tau^2r^2}{(\tau^2+r^2)^2}\right).
	\eea
Then, the two different linear combinations corresponding to the contribution of the $F^2$ operator verses the $F^3$ operator vanish for different values of $\tau$ and $r$. For example, for $d=3$ Euclidean dimensions, 
	\bea
		\left\langle J^a_1(x)J^b_1(y)J^c_2(z) \right\rangle &=& \frac{f^{abc}}{2^9g_{SG}^2}\frac{1}{\big[ 2r(\tau^2+r^2)\big]^4}\left((3r^4-2\tau^4+13\tau^2r^2)-\frac{2}{\Delta^2}(r^4-2\tau^4+19\tau^2r^2) \right) \nonumber
	\eea
Comparing the to measurements at the two different set of points where either contribution vanishes, we can not only test the validity of the effective approach, but also find the dimension gap $\Delta$ suppressing higher order corrections.

To conclude, in this paper we computed the three-point current correlation function in the framework of Effective Conformal Field Theory. This describes the dynamics of all operators with dimensions below the cutoff dimension $\Delta_{Heavy}$. In systems with large dimension gap $\Delta_{gap}\approx \Delta_{heavy}$ and a $1/N$ like suppression, there is double expansion in both $1/N$ and $1/\Delta_{gap}$. The contributions to the three-point current correlation function coming from the lowest non-renormalizable bulk operator $F^3$ is computed and compared to the contribution coming from the renormalizable $F^2$ bulk operator already computed in the literature. It is shown that the two operators give rise to different polarization structure of the three-point current correlation function. The polarization structure coming from the non-renormalizable bulk $F^3$ term is suppressed by powers of the cutoff dimension $\Delta_{Heavy}$ prescribed by demanding perturbative unitarity. 

By measuring the non-linear response to external fields, it is possible to test the effective description for strongly coupled condensed matter systems. In systems with global non-Abelian symmetry and large hierarchy in operator dimensions at second order phase transition, we can expect new terms of order $1/\Delta_{Heavy}^2$ in the three-point current correlation function with a different polarization structure to the leading effect. 

%

\section*{Acknowledgment} 

This humble effort would not have been possible without the help of persons whom I would like to acknowledge in this section. First and foremost I thank God to whom this work, the first fruit of my labor in graduate school, is dedicated. 
I thank Jared Kaplan for suggesting the project and for many helpful discussions, both physics related and otherwise. I thank Michael Peskin for being very supportive of my efforts. I also thank Srinivas Raghu and Maissam Barkeshli for helpful conversations about the experimental aspects of the work, and Laim Fitzpatrick for helpful feedback. The work was supported by DOE under contract DE--AC02--76SF00515.

\appendix
\section{Calculation of $\mathcal{F}^{(3)}_{ijk}$}
We begin with (\ref{Fijk2}).
	\bea
		\mathcal{F}^{(3)}_{ijk}(0,y,z)&=&(d-2)^3(C^d)^3|y|^{2(d-1)}J_{aj}(y')|z|^{2(d-1)}J_{bk}(z) \nonumber\\
		&& \int d^dw'dw'_0\frac{(w'_0)^{2d-4}}{[w'^2_0+(\tilde{x}'-y')^2]^{d-1}[w'^2_0+(\tilde{x}'-z')^2]^{d-1}}\nonumber\\
		&&\hspace{1.7cm} \bigg(J_{0[i}(w'-y')J_{\gamma]a}(w'-y')J_{0[\gamma}(w'-z')J_{0]b}(w'-z')\nonumber\\
		&&\hspace{1.8cm}   +J_{0[\gamma}(w'-y')J_{0]a}(w'-y')J_{0[\gamma}(w'-z')J_{i]b}(w'-z')\bigg)
	\eea
The following integral appears repeatedly in the evaluation of $\mathcal{F}^{(3)}_{ijk}.$ It was performed using Feynman parameters in \cite{Freedman:1998tz}. In the following $x, y, z, w$ are coordinates in the $d$-dimensional boundary, and $z_0, w_0$ are perpendicular bulk coordinates.
	\bea
	\label{Iabcd}
		\int_{0}^{\infty}dz_0 \int d^dz\frac{z_0^a}{[z_0^2+(z-x)^2]^b[z_0^2+(z-y)^2]^c}
		&\equiv& I[a,b,c,d]|x-y|^{1+a+d-2b-2c}\\
		I[a,b,c,d]
		&=&   \frac{\pi^{\frac{d}{2}}}{2}\frac{\Gamma \left( \frac{a+1}{2}\right)\Gamma \left(b+c-\frac{d}{2}-\frac{a+1}{2} \right)}
				{\Gamma(b)\Gamma(c)}\nonumber\\
		&&\frac{\Gamma \left(\frac{a+1}{2}+\frac{d}{2}-b\right)\Gamma \left( \frac{a+1}{2}+\frac{d}{2}-c\right)}{\Gamma(a+1+d-b-c),}
	\eea
We can proceed in the evaluation of $\mathcal{F}^{(3)}_{ijk}$ by expressing the tensors in the integrand in terms of derivatives of the integrand in the left hand side of (\ref{Iabcd}) as follows:
	\bea
		\frac{J_{kl}(w-t)}{[w^2_0+(w-t)^2]^{d-1}}
		&=& \left(\frac{d}{d-1} \right)\frac{\delta_{kl}}{[w^2_0+(w-t)^2]^{d-1}}
			-\frac{1}{2(d-1)(d-2)}\frac{\p}{\p t_k}\frac{\p}{\p t_l}\left( \frac{1}{[w^2_0+(w-t)^2]^{d-2}} \right)\hspace{5mm}\\
			\
		\frac{(w-t)_j(w-t)_i}{[w^2_0+(w-t)^2]^{d}}
		&=& \frac{1}{2(d-1)}\frac{\delta_{ji}}{[w^2_0+(w-t)^2]^{d-1}}
			+\frac{1}{4(d-2)(d-1)}\frac{\p}{\p t_j}\frac{\p}{\p t_i}\left( \frac{1}{[w^2_0+(w-t)^2]^{d-2}} \right)\\
			\
		\frac{J_{i[j}(w-t)J_{k]l}(w-t)}{[w^2_0+(w-t)^2]^{d-1}}
		&=& \frac{\delta_{i[j}\delta_{k]l}}{[w^2_0+(w-t)^2]^{d-1}}\nonumber\\
			&&-\frac{1}{2(d-1)(d-2)}\left( \delta_{i[j}\p^t_{k]} \p^t_l + \delta_{l[k}\p^t_{j]}\p^t_i\right)
			\left( \frac{1}{[w^2_0+(w-t)^2]^{d-1}} \right)
	\eea
where, 
	\begin{eqnarray*}
		t
		&=& (y-x)'-(z-x)' = \frac{y-x}{|y-x|^2}-\frac{z-x}{|z-x|^2}\\
		\p^t_k &=& \frac{\p}{\p t_k}
	\end{eqnarray*}

The integral on the right hand side of Eq.(\ref{Fijk2}) now simplifies to the following.
	\[
		\int d^dw'dw'_0\: (w'_0)^{2d-4}\bigg( A + B + C-A'-B'-C'\bigg)
	\]

	\bea
	A&=& -\frac{2}{d-1}\delta_{a[b}\p^{y'}_{i]}
		\frac{w'^3_0}{[w'^2_0+(\vec{w'}-y')^2]^{d-1}[w'^2_0+(\vec{w'}-z')^2]^{d}}\nonumber\\
	B&=& -\frac{1}{d-1}\delta_{a[i}\p^{y'}_{b]}
		\frac{w'_0}{[w'^2_0+(\vec{w'}-y')^2]^{d-1}[w'^2_0+(\vec{w'}-z')^2]^{d-1}}\nonumber\\
	C&=& -\frac{1}{2(d-2)(d-1)}\bigg(\p^{y'}_i\p^{z'}_a\p^{z'}_b - \delta_{ai}\p^{y'}_c\p^{z'}_b\p^{z'}_c\bigg)
		\frac{w'_0}{[w'^2_0+(\vec{w'}-y')^2]^{d-1}[w'^2_0+(\vec{w'}-z')^2]^{d-2}}
	\eea
$A',$ $B',$ and $C'$ are just $A,$ $B,$ and $C$ with the substitutions $y' \leftrightarrow z',$ and $a \leftrightarrow b$.

We find the following results for the integrals
	\bea
		\int d^dw'dw'_0\: (w'_0)^{2d-4}\big( A \big)
		&=& \pi^{d/2}\frac{\left[ \Gamma \left( \frac{d}{2} \right )\right]^3}{\left[ \Gamma(d) \right]^2}
			\frac{\delta_{a[i}(y'-z')_{b]}}{|y'-z'|^d}\\
		\
		\int d^dw'dw'_0\: (w'_0)^{2d-4}\big(B \big)
		&=& \pi^{d/2}\frac{\left[ \Gamma \left( \frac{d}{2} \right )\right]^3}{\left[ \Gamma(d) \right]^2}
			\frac{\delta_{a[i}(y'-z')_{b]}}{|y'-z'|^d}\\
		\
		\int d^dw'dw'_0\: (w'_0)^{2d-4}\big(C \big)
		&=& -\frac{\pi^{d/2}}{2(d-1)}\frac{\left[ \Gamma \left( \frac{d}{2} \right )\right]^3}{\left[ \Gamma(d) \right]^2}
			\frac{1}{|y'-z'|^d} \Bigg ( \delta_{ab}(y'-z')_i + \delta_{ib}(y'-z')_a - \delta_{ai}(y'-z')_b \nonumber\\
			&& \hspace{1.9in}-\frac{d}{|y'-z'|^2}(y'-z')_i(y'-z')_a(y'-z')_b \Bigg )
	\eea
Therefore, putting all of the pieces together, we find
	\bea
		\int d^dw'dw'_0\: (w'_0)^{2d-4}\bigg( A + B + C-A'-B'-C'\bigg)  \nonumber\\
		= \frac{\pi^{d/2}}{d-1}\frac{\left[ \Gamma \left( \frac{d}{2} \right )\right]^3}{\left[ \Gamma(d) \right]^2}
			\frac{1}{|y'-z'|^d} \Bigg( -\delta_{ab}(y'-z')_i + \frac{d}{|y'-z'|^2}(y'-z')_i(y'-z')_a(y'-z')_b \Bigg)
	\eea
Which gives the following result for $\mathcal{F}^{(3)}_{ijk}(0,y,z)$
	\bea
		\mathcal{F}^{(3)}_{ijk}(0,y,z)
		&=& \kappa(d-2)^2
		\frac{1}{|y|^{2(d-1)}} J_{aj}(y) \frac{1}{|z|^{2(d-1)}} J_{bk}(z)
		\frac{1}{|t|^d}\left( -\delta_{ab}t_i + \frac{d}{|t|^2}t_it_at_b \right) \qquad\\
		\kappa &=& \pi^{d/2}(C^d)^3\frac{(d-2)}{(d-1)}\frac{\big[ \Gamma(\frac{d}{2})\big]^3}{\big[ \Gamma(d)\big]^2}, \quad C^d = \frac{\Gamma(d)}{2\pi^{d/2}\Gamma(\frac{d}{2})}\nonumber
	\eea
To restore the $x$ dependence we make the replacements $y \rightarrow y-x$ and $z \rightarrow z-x$\\ and find $\mathcal{F}^{(3)}_{ijk}(0, y-x, z-x).$ This is related to $\mathcal{F}^{(3)}_{ijk}(x, y, z)$ by shift symmetry. \\Using $t^2 = (y-z)^2/\left[ (z-x)^2(y-x)^2\right]$, we find,
	\bea
		\mathcal{F}^{(3)}_{ijk}(0, y-x, z-x)
		&=&\kappa(d-2)^2 \frac{J_{lj}(y-x)}{|y-x|^{d-2}}\frac{J_{mk}(z-x)}{|z-x|^{d-2}}
			\frac{1}{|z-y|^d}\left( -\delta_{lm}t_i +\frac{d}{|t|^2}t_it_lt_m \right).
	\eea
Finally, we can express $\mathcal{F}^{(3)}_{ijk}$ in terms of $C_{ijk}$ and $D_{ijk}$ in the following manner:
\bea
	\frac{t_lt_m}{t^2}&=&-\frac{1}{2}\left( J_{lm}(t) - \delta_{lm} \right)\nonumber\\
	J_{lj}(y-x)J_{lm}(t)J_{mk}(z-x)&=&J_{lj}((y-x)')J_{lm}((y-x)'-(z-x)')J_{mk}((z-x)')\nonumber\\
	&=& J_{jk}(y-z)
\eea
And we arrive at the following final expression for $\mathcal{F}^{(3)}_{ijk}$:
	\bea
		\mathcal{F}_{ijk}(x,y,z) &=& -\kappa\frac{d}{2}\left(D_{jki}(y,z,x) + \frac{1}{d}C_{jki}(y,z,x)\right), \nonumber \\
	\eea
	

%

\end{document}